\documentclass{article}

\usepackage{arxiv}
\usepackage[utf8]{inputenc} % allow utf-8 input
\usepackage[T1]{fontenc}    % use 8-bit T1 fonts
\usepackage{hyperref}       % hyperlinks
\usepackage{url}            % simple URL typesetting
\usepackage{booktabs}       % professional-quality tables
\usepackage{amsfonts}       % blackboard math symbols
\usepackage{nicefrac}       % compact symbols for 1/2, etc.
\usepackage{microtype}      % microtypography
\usepackage{lipsum}
\usepackage{graphicx}
\usepackage{amsmath}
\usepackage{CJKutf8} 
\usepackage{float}

\graphicspath{ {./images/} }

\title{\textbf{CAS-Canglong: A skillful 3D Transformer model for sub-seasonal to seasonal global sea surface temperature prediction} }

\author{
Longhao Wang\textsuperscript{†} \\
  IGSNRR\\
  Chinese Academy of Sciences\\
  Beijing, China\\
  \texttt{wanglonghao0857@isgnrr.ac.cn}
  \And
 Xuanze Zhang\textsuperscript{†} \\
  IGSNRR\\
  Chinese Academy of Sciences\\
  Beijing, China \\
  \texttt{xuanzezhang@igsnrr.ac.cn} \\
  \And
 L. Ruby Leung \\
  Pacific Northwest National Laboratory\\
  Richland, Washington 99352, USA\\
  \texttt{ruby.leung@pnnl.gov} \\
  \And
 Francis H.S. Chiew \\
  CSIRO Environment\\
  Black Mountain, Canberra, ACT 2601, Australia\\
  \texttt{francis.chiew@csiro.au} \\
  \And
 Amir AghaKouchak \\
  Center for Hydrometeorology and Remote Sensing\\
  University of California, Irvine\\
  California, USA\\
  \texttt{amir.a@uci.edu} \\
  \And
 Kairan Ying \\
  National Institute of Natural Hazards\\
  Ministry of Emergency Management of China\\
  Beijing, China\\
  \texttt{yingkr@tea.ac.cn} \\
  \And
 Yongqiang Zhang* \\
  IGSNRR\\
  Chinese Academy of Sciences\\
  Beijing, China \\
  \texttt{zhangyq@igsnrr.ac.cn} \\
}

\begin{document}

\maketitle
\begingroup
\renewcommand\thefootnote{\fnsymbol{footnote}}

\footnotetext[2]{These authors contributed equally to this work.}
\endgroup

\begin{abstract}
Accurate prediction of global sea surface temperature at sub-seasonal to seasonal (S2S) timescale is critical for drought and flood forecasting, as well as for improving disaster preparedness in human society\textsuperscript{ \cite{RN1,RN2,RN3}}. Government departments or academic studies normally use physics-based numerical models to predict S2S sea surface temperature and corresponding climate indices, such as El Niño-Southern Oscillation. However, these models are hampered by computational inefficiencies, limited retention of ocean-atmosphere initial conditions, and significant uncertainty and biases\textsuperscript{\cite{RN4,RN5}}. Here, we introduce a novel three-dimensional deep learning neural network to model the nonlinear and complex coupled atmosphere-ocean weather systems. This model incorporates climatic and temporal features and employs a self-attention mechanism to enhance the prediction of global S2S sea surface temperature pattern. Compared to the physics-based models, it shows significant computational efficiency and predictive capability, improving one to three months sea surface temperature predictive skill by 13.7 ± 3.0\% to 77.1 ± 0.6\% in seven ocean regions with dominant influence on S2S variability over land. This achievement underscores the significant potential of deep learning for largely improving forecasting skills at the S2S scale over land.
 
\end{abstract}

% keywords can be removed

\section{Introduction}
At timescales ranging from two weeks to a few months, S2S prediction is a major challenge at the weather-climate nexus\textsuperscript{\cite{RN6,RN7}}. Despite steady progress in the last two decades, prediction skill at the S2S scale remains very low\textsuperscript{\cite{RN8,RN9,RN10,RN11}}. Accurate S2S prediction of global sea surface temperature (SST), as one of the key initial and boundary conditions of numerical weather and climate prediction models, is crucial for improving the predictive skill of S2S weather and climate events, due to the strong teleconnections of global SST anomaly and associated modes of climate variability to extreme events through modulation of atmospheric circulations\textsuperscript{\cite{RN12}}. For example, it is well-known that the El Niño-Southern Oscillation (ENSO) featuring tropical SST anomaly in the tropical Pacific Ocean plays a central role in seasonal to annual regional climate prediction\textsuperscript{\cite{RN13,RN14,RN15}}. Consequently, S2S prediction of ENSO events can provide early warning of extreme events over ENSO-influenced regions with lead times ranging from several weeks to a few months\textsuperscript{\cite{RN16,RN17}}. With anthropogenic warming, extreme weather and climate events, such as record-breaking heat waves, flash or years-long droughts, torrential rains, severe floods, and coastal inundation from tropical cyclones are becoming more frequent and more severe\textsuperscript{\cite{RN1,RN18,RN19}}, causing significant loss of life and economic damage. Therefore, developing robust S2S prediction methods is critical to meet the increasing demand for extreme events forecasting, early warning, and disaster preparedness. 

Numerical modelling has been widely used by international major forecasting centres for predicting S2S variations of global SST\textsuperscript{\cite{RN20,RN21}}. However, accurately predicting the S2S SST variability using general circulation models (GCMs) remains a great challenge, particularly for extratropical SST because GCMs exhibit large biases owing to inadequate representation of sea-air interactions and atmospheric processes connected to large-scale atmospheric circulations\textsuperscript{\cite{RN22}}. Utilizing partial differential equations and physics parameterizations to represent the coupled ocean-atmosphere system, the large computational requirements of GCMs have limited their spatial resolution and ensemble size, contributing to their relatively low S2S SST prediction skill\textsuperscript{\cite{RN5,RN23}}. As the predictability of the atmosphere diminishes rapidly after one month, uncertainty in predicting the global SST pattern plays an important role in limiting predictability at the S2S timescale\textsuperscript{\cite{RN4,RN24,RN25}}.

Unlike physics-based modelling methods, data-driven deep learning (DL) models leverage high-dimensional and long-time-scale multivariate data can be used to predict SST by learning patterns and intrinsic physical relationships from historical observations. Previous studies have used the Long Short-Term Memory (LSTM) network in SST prediction\textsuperscript{\cite{RN6,RN26,RN27,RN28}}. Different variants of the LSTM methods have become the de facto standard due to their ability to capture contextual information and suitability for time-series SST modelling. However, LSTMs primarily focus on local pixel data, effectiveness their application in high-dimensional multivariate modelling and failing to capture the complex dynamics of ocean-atmosphere interactions\textsuperscript{\cite{RN29}}. This limitation renders them reduces their efficacy for sub-seasonal weather prediction and detecting extreme weather anomalies.

Recent advances in DL-based artificial intelligence (AI) models have spurred their applications in the Earth science community, particularly in short to medium-range global weather forecasting\textsuperscript{\cite{RN30,RN31,RN32,RN33}}, and seasonal-to-annual ENSO prediction\textsuperscript{\cite{RN34,RN35}}. However, to date, there has been no reported use of DL models specifically S2S global SST prediction. Here we develop an AI-based model using a self-attention-based Transformer architecture, named as CAS-Canglong for predicting SST at S2S timescale. Our implementation of the three-dimensional (3D) Transformer establishes multivariate relationships across space and time\textsuperscript{\cite{RN35}}. Specifically, we use 51 years (1959-2009) of reanalyses data for eight key oceanic and atmospheric variables to train the CAS-Canglong, while validating and testing the model in 2009-2015 and 2016-2022, respectively. Results show that CAS-Canglong demonstrates superior and robust capabilities in predicting SST seasonal cycles, anomalies, and SST-based climate indices in both hindcast and near-real-time forecasting modes, as detailed in the Result Section.

\section{Result}
\subsection{Hindcast comparison}
\begin{figure}[H] % picture
    \centering
    \includegraphics[width=1\columnwidth]{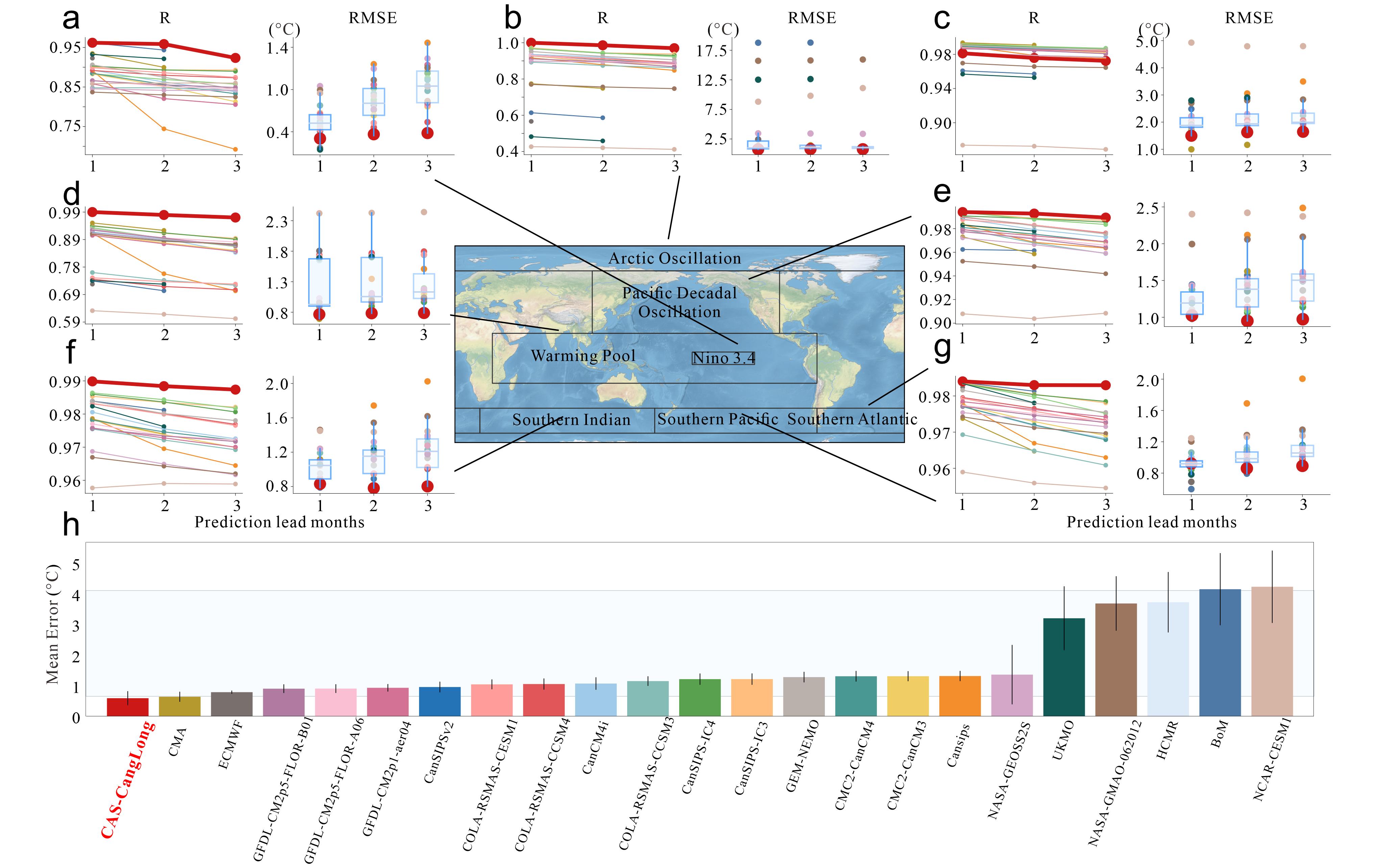}
    \caption{\textbf{Predictive skill assessment for the deep learning-based S2S model (CAS-Canglong) and the state-of-the art GCM S2S models. }The one to three-month lead predictions of SST from CAS-Canglong are compared to those from 22 state-of-the-art NWP models. The correlation skill (line plot) and latitude-weighted RMSE (point and box plot) are summarized, by key SST regions and associated SST climate indices. For a robust comparison, the calculated correlation skill and RMSE (in ℃) are based on the mean values from 2015 to 2020, which represents an overlap between CAS-Canglong's testing period and availability of the physical-based data. The boxplot shows the RMSE distribution of different models. Each sub-boxplot shows the lower whisker (Q1-1.5IQR), lower quartile (Q1), median, upper quartile (Q3), and upper whisker (Q3+1.5IQR), where the IQR = Q3-Q1. The regions or climate indices are as follows: (\textbf{a}) Niño 3.4 region (170°W-120°W, 5°S-5°N). (\textbf{b}) Arctic Oscillation region (75°N-90°N). (\textbf{c}) Pacific Decadal Oscillation (PDO) region, defined by the leading pattern (EOF) of SSTA in the North Pacific (110°E-100°W, 20°N-75°N). (\textbf{d}) tropical Warming Pool (WP) region (35°E-75°W, 15°S-15°N). (\textbf{e}) Southern Atlantic Ocean (SAO) region (20°E-160°E, 40°S-60°S). (\textbf{f}) Southern Indian Ocean (SIO) region (160°E-110°W, 40°S-60°S), and (\textbf{g}) Southern Pacific Ocean (SPO) region (20°E-160°E, 40°S-60°S). (\textbf{h}) Mean error (in ℃) of the CAS-Canglong and the numerical S2S models compared with the ERA5 ground truth. In \textbf{h}, all predictions are for one-month lead of global SST from 2015 to 2020. The shaded areas denote the 5\textsuperscript{th} to 95\textsuperscript{th} percentile of ensemble model prediction. The error bars represent the mean error ± one standard deviation. Model colors in (\textbf{a})-(\textbf{g}) correspond to those in (\textbf{h}), where the model names are shown. }
    \label{fig:fig1}
\end{figure}

To quantitatively assess the performance of our AI-model’s SST predictions, we compared CAS-Canglong with the numerical models based predictions obtained from ECMWF S2S model ensemble\textsuperscript{\cite{RN36}} and NMME North American multi-model ensemble\textsuperscript{\cite{RN37}} for the overlapping period 2015-2020. The CAS-Canglong model has a spatial resolution of 0.25°×0.25°, which is the finest available and comparable to that of the physics-based models. The spatial resolution of our model remapped to match the resolution of the physics-based models, most of which have a spatial resolution of 1°×1°. The prediction lead time is set to 3 months for the assessment, and we compare the correlation (R) and Root Mean Square Error (RMSE) of SST predictions in several key regions and associated climate indices. The predictive skill for SST across different regions are summarised in Fig \ref{fig:fig1}, indicating that the CAS-Canglong model gives better SST predictions for one, two and three months lead time compared to the numerical models. From Fig. \ref{fig:fig1}, CAS-Canglong shows more accurate and reliable predictions than the dynamical models from one- to three-month scales (see also Fig. S1). The CAS-Canglong model performs best for both the correlation skill in six of the seven regions (except for the Pacific Decadal Oscillation). Comparing the prediction error with all the GCMs, CAS-Canglong achieves the lowest error, surpassing all the numerical models. The improvement (reduction of RMSE) at the S2S scale (i.e., one to three months mean) is 34.9 ± 3.3\%, 77.1 ± 0.6\%, 28.5 ± 3.7\%, 39.6 ± 1.0\%, 31.2 ± 2.7\%, 30.3 ± 2.1\%, and 13.7 ± 3.0\%, for the regions of Niño 3.4, Artic Oscillation, Pacific Decadal Oscillation, Warm Pool, Southern Atlantic Ocean, Southern India Ocean, and Southern Pacific Ocean, respectively. For a robust comparison, we also evaluated our SST predictions against the Hadley Centre Sea Surface Temperature data set. Using the same evaluation method, CAS-Canglong again achieved outstanding performance for all the regions investigated (Fig. S2). The spatial distribution and correlations for the global SST and SST anomalies forecasts are illustrated in Fig. S3 and Fig. S4. Notably, our model generates SST predictions for lead times ranging from 1 to 24 months, with remarkably low prediction errors. Additionally, CAS-Canglong model nearly matches the current state-of-the-art SST prediction models, producing predictions highly consistent with those of the ECMWF (Fig. S5). Based on the outstanding overall performance of CAS-Canglong, we focus on its results in key SST regions compared with the multi-models in the following sections. 
\subsection{Hindcast in major regions}
\begin{figure}[H] % picture
    \centering
    \includegraphics[width=1.0\columnwidth]{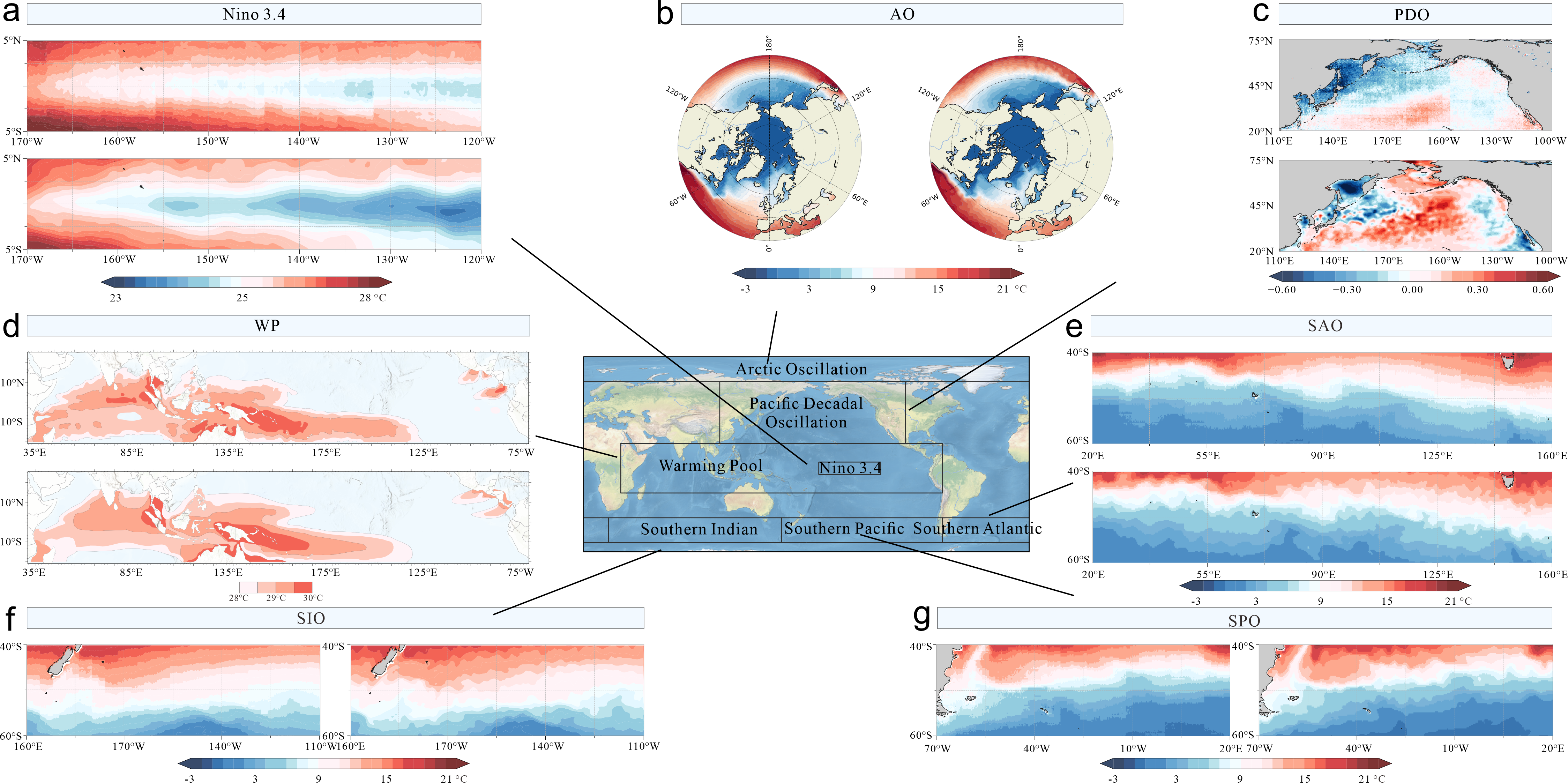}
    \caption{\textbf{Prediction of SST and related climate indices in key global regions. }An example of spatial map in March 2012 with a weak La Niña event. For each panel of a specific SST region, the spatial maps at the top or on the left show the three-month lead predictions of CAS-Canglong, with the corresponding spatial maps at the bottom or on the right showing the ERA5 ground truth. Corresponding with \textbf{Fig. \ref{fig:fig1}}, the regions are for (\textbf{a}) Niño 3.4 (\textbf{b}) AO, (\textbf{c}) PDO, (\textbf{d}) WP, (\textbf{e}) SIO, (\textbf{f}) SPO, and (\textbf{g}) SAO. In \textbf{d}, the warm pool boundary contour lines are rendered smoothly to highlight the model's accuracy in capturing the extent of the warm pool.  }
    \label{fig:fig2}
\end{figure}

\begin{figure}[H] % picture
    \centering
    \includegraphics[width=1.0\columnwidth]{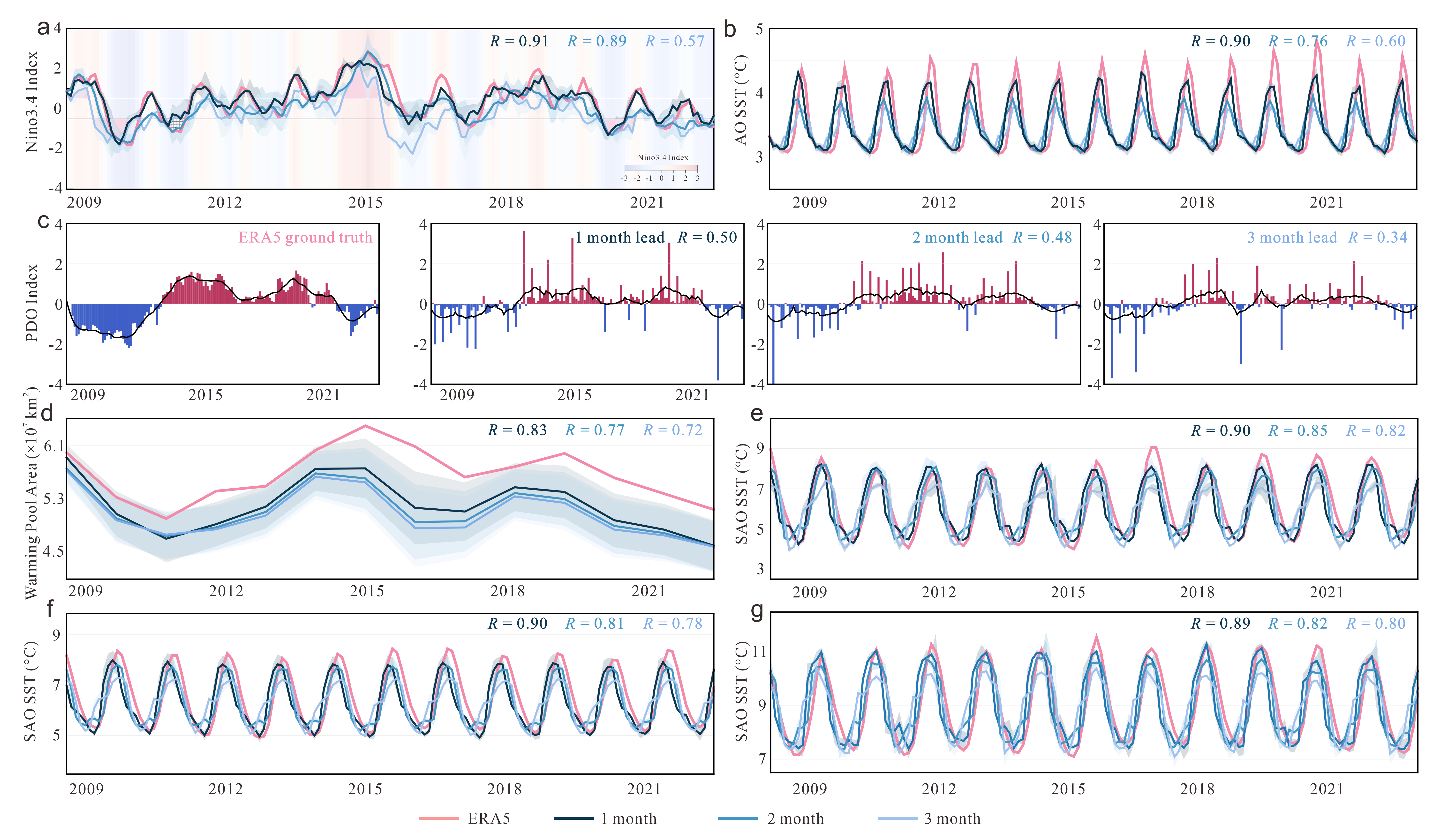}
    \caption{\textbf{Time series of SST predictions and climate indices in seven major regions for the period 2009-2022. }The red line represents the ERA5 ground truth, while the blue lines show the model predictions for one to three months lead time, with colors ranging from dark to light blue corresponding to one, two, and three months, respectively. The background shading of the time series is overlaid with ± one standard deviation. The regions correspond with Fig. \ref{fig:fig1} \& Fig. \ref{fig:fig1}: (\textbf{a}) Niño 3.4. (\textbf{b}) AO. (\textbf{c}) PDO. (\textbf{d}) WP. (\textbf{e}) SIO. (\textbf{f}) SPO, and (\textbf{g}) SAO. The warm pool boundary contour lines are rendered smoothly to highlight the model's accuracy in capturing the extent of the warm pool. The Niño index (shaded) is marked along the x-axis in (\textbf{a}).
}
    \label{fig:fig3}
\end{figure}

Here, we focus on the seven SST regions and associated indices (see Methods), including the regions of Niño 3.4, Artic Oscillation (AO), Pacific Decadal Oscillation (PDO), warm pool (WP), Indian Ocean Dipole (IOD), Southern Atlantic Ocean (SAO), Southern Indian Ocean (SIO), and Southern Pacific Ocean (SPO). Fig. \ref{fig:fig2} shows the 3-month hindcast SST or SST-based climatic indices for these regions versus the ERA5 ground truth. The CAS-Canglong model shows strong forecast skill in all the regions. In tropical regions, specifically the Indo-Pacific warm pool area with SSTs exceeding 28°C, the model's performance is shown during the single-month of March 2012 with a weak La Niña event (Fig. \ref{fig:fig2}) and across the multi-year mean warm pool area (Fig. S6). The predicted warm pool area is 5.93×10\textsuperscript{7} km², compared to the ERA5 of 6.17×10\textsuperscript{7} km², resulting in a low error of 3.8\% (Fig. \ref{fig:fig2}d). Fig. S7 further shows the predicted Indian Ocean Dipole (IOD) compared with the ERA5 ground truth, exhibiting a high correlation skill (R = 0.63). For the temperate Southern Oceans, including the SIO, SPO, and SAO (Figs. \ref{fig:fig2}e-g and \ref{fig:fig3}e-g), the spatial distribution of predicted SSTs is highly consistent with the ERA5. For the three-month lead predictions, the model shows superior performance for the SIO, SPO, and SAO with a correlation value greater than 0.8 for all Southern Ocean regions (Fig. \ref{fig:fig3}e-g). In boreal high latitudes, our model successfully captures the spatial distribution of Arctic SSTs (Fig. S8). These results indicate our model’s strong performance not only in tropical but also in temperate and high latitude regions. 

We now focus on the entire test period from 2009 to 2022 for the SST regions and related climate indices. The CAS-Canglong time series and the ground truth for the SST and related climate indices are presented in Fig. \ref{fig:fig3}, with predictions categorized by lead times of 1 to 3 months. We select the most important SST index at S2S timescale, the ENSO-related Niño index, to elucidate the model's predictive capabilities. Our analysis emphasizes the accuracy in forecasting El Niño and La Niña events across the entire test dataset. The temporal evolution of the predicted Niño 3.4 index closely aligns with the ERA5 for one- to three-month lead predictions, achieving correlation values of 0.91, 0.89, and 0.57, respectively (Fig. \ref{fig:fig3}a). The CAS-Canglong surpasses the current state-of-the-art model ECMWF, with higher correlation and prediction skill (Fig. S9). The model accurately captures all the ENSO events during 2009-2022, achieving an overall hit rate of 85\% for three months in advance. In the tropical region, our model demonstrates outstanding performance not only in simulating the Niño 3.4 index but also the Warm Pool area. As shown in Fig. \ref{fig:fig3}d, the model accurately captures the variations of the dynamic Warm Pool area, with correlation skill > 0.7 for three months in advance. In the temperate regions, our model successfully tracks the phase transitions in the PDO index (Fig. \ref{fig:fig3}c). In the Southern Ocean, the model shows its superior performance, with prediction correlations reaching 0.9 and 0.7 for one-month and three-month lead time, respectively. The model well captures the seasonal variations in SST of the Southern Ocean. For the AO in polar regions, where traditional NWP models often struggle to simulate and predict, our model also exhibits remarkable performance, with SST prediction correlations reaching 0.9 and 0.6 for one-month and three-month lead times, respectively (Fig. \ref{fig:fig3}b). 
\subsection{Near real-time forecasting}
\begin{figure}[H] % picture
    \centering
    \includegraphics[width=1.0\columnwidth]{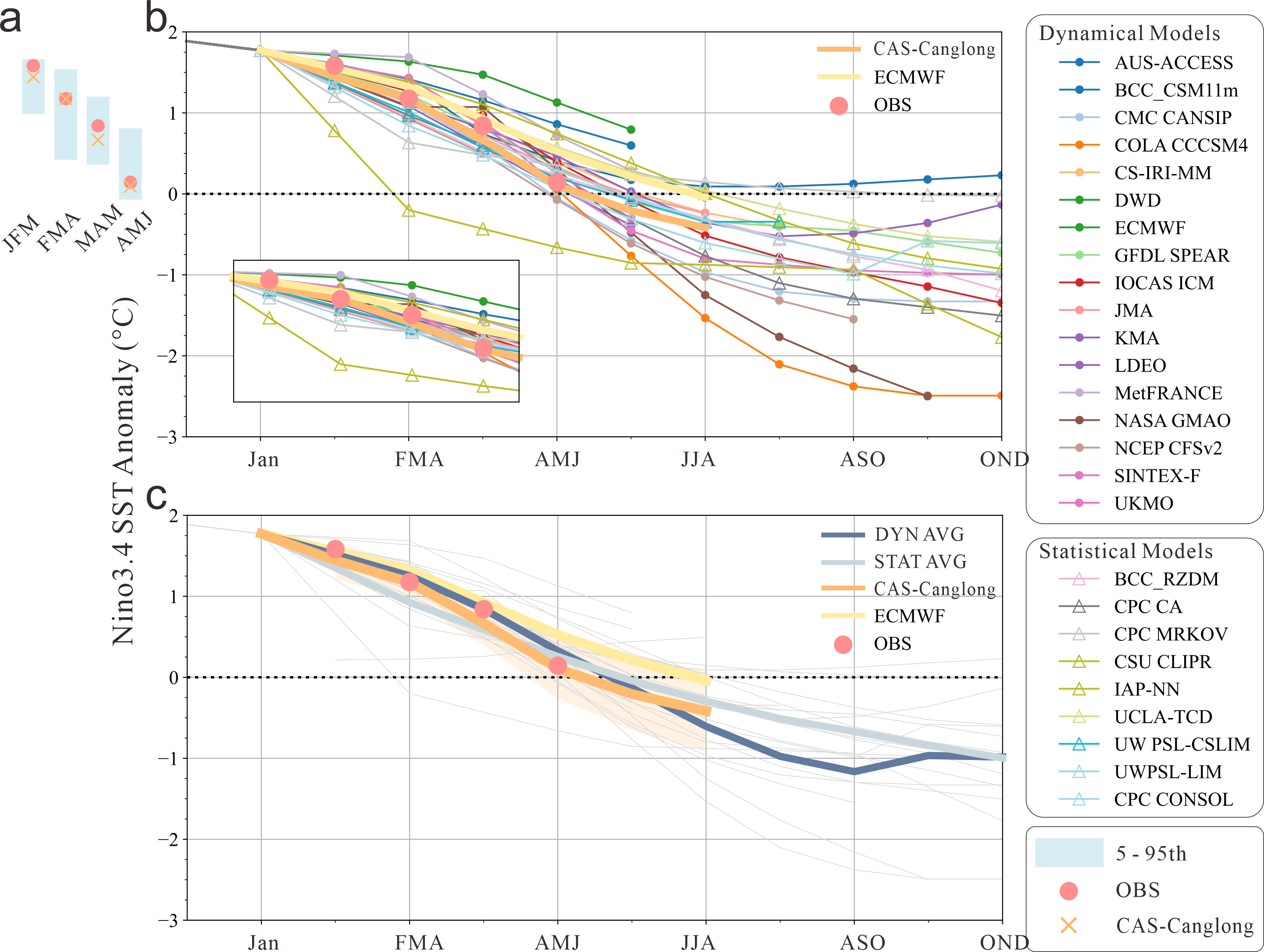}
    \caption{\textbf{Near real-time predictions for the Niño 3.4 SST anomalies since January 2024.} These ensemble model datasets were obtained in January 2024 from the International Research Institute (IRI), with the ERA5 ground truth updated in real-time. (\textbf{a}) The magnified section of the graph shows the data from January to May 2024. The blue boxes indicate the 95th percentile of the ensemble predictions from all dynamical and statistical models, yellow crosses represent CAS-Canglong’s predictions, and red dots mark the ERA5 ground truth, highlighting CAS-Canglong’s prediction within the ensemble predictions of all other models and, its proximity to the true values. (\textbf{b}) Multi-model predictions for the Niño 3.4 SST anomalies are displayed, including our model's predictions alongside the ERA5 ground truth (red dot). (\textbf{c}) Other model lines are faded out and, summarized as IRI statistical ensemble-averaged predictions (dark blue) and dynamical ensemble-averaged forecasts (blue). Lines for the ECMWF S2S model and the CAS-Canglong are highlighted for comparison. 
}
    \label{fig:fig4}
\end{figure}

To evaluate our model's real-time forecasting performance, Fig. \ref{fig:fig4} presents the predicted SST anomalies for the Niño 3.4 region in 2024, compared against data for March to June 2024 obtained from the latest ERA5 observations. Availability of the latter allows us to forecast the subsequent months (July-December). The ensemble models’ predictions from the International Research Institute (IRI) are also shown in Fig. \ref{fig:fig4} In the first three months of the year 2024, a typical El Niño event with positive SST anomaly is observed, with a potential shift towards a La Niña event with negative SST anomaly in the latter half of the year. The CAS-Canglong model successfully predicts this transition, with higher predictive accuracy than the ECMWF S2S model. Fig. \ref{fig:fig4} also demonstrates that during the El Niño months, our model achieves robust results, nearly matching the ensemble mean of the current state-of-the-art dynamical models or statistical models. The CAS-Canglong model captures the observed earlier transition time point (during AMJ) than both the ensemble predictions of dynamical and statistical models These results validate our model's capability to effectively perform S2S scale predictions in near real time, offering both faster inference speeds and enhanced accuracy.
\section{Discussion}
Forecasting at the S2S scale across global land is critical for early warning and emergency preparedness of major extreme weather and climate events such as heatwave and droughts. The seasonal variability of SST on the S2S scale has profound impacts on atmospheric variability over land so accurate prediction of SST is necessary for predicting extreme events that affect terrestrial ecosystems and human societies\textsuperscript{\cite{RN38}}. However, traditional physics-based dynamical models still have low predictability on the S2S scale\textsuperscript{\cite{RN39}}. The major challenge is that the S2S time range is much longer than the memory of atmospheric processed but is too short for the sea surface variability to have a strong influence relative to the variability of atmospheric circilations\textsuperscript{\cite{RN4,RN7}}. Global sea surface temperatures (SSTs), particularly the tropical SSTs over the Indo-Pacific Warm Pool region and the eastern tropical Pacific Ocean (e.g., the Niño 3.4 region), exert the dominant roles in global subseasonal-to-seasonal variability of hydroclimate extremes by releasing huge amounts of water vapour and latent heat to the atmosphere which modulate atmospheric circulations and the spatiotemporal distributions of global land precipitation\textsuperscript{\cite{RN40}}. SST is a crucial factor in predicting precipitation and extreme weather events and assessing extreme weather risks\textsuperscript{\cite{RN40}}. Furthermore, ENSO events on the S2S scale strongly impact extreme precipitation and droughts, with outsized impacts in densely populated regions\textsuperscript{\cite{RN41}}. Therefore, enhancing the accuracy of SST predictions on the S2S timescales could substantially improve forecasting of extreme events which are increasingly affecting humans and societies due to anthropogenic warming.

Recent advances in data-driven, attention mechanism-based deep learning models offer a promising approach for weather forecasting and seasonal-to-annual ENSO predictions\textsuperscript{\cite{RN35,RN42}}. Inspired by the recent successes of the Transformer model in computer vision\textsuperscript{\ref{RN43,RN44}}, such as the Sora model's ability to predict video sequences up to one minute long\textsuperscript{\cite{RN45}}, we recognized the potential for similar applications in weather forecasting and climate prediction for enhancing long-term predictive capabilities in Earth sciences\textsuperscript{\cite{RN46}}. Consequently, we developed the CAS-Canglong model utilizing the Swin-Transformer architecture and unique self-attention mechanisms to advance long-term modeling and prediction capabilities. Our model demonstrates strong performance in sub-seasonal SST predictions, exhibiting high correlation skills, with the potential to improve S2S predictions of weather and climate anomalies. The CAS-Canglong model represents a breakthrough, successfully simulating and predicting global SST on sub-seasonal to seasonal scales for the first time using deep learning.

CAS-Canglong may enable early warning of extreme events by effectively predicting SST on sub-seasonal to seasonal scales, capturing both the seasonal cycles and anomalies in sea surface temperatures. The predictive prowess of our model has been demonstrated, not only for ENSO events but also for other modes of variability such as the Tropical Warm Pool area, AO, PDO, and the Southern Ocean SST anomalies, demonstrating a profound ability to model the air-sea interactions processes. This understanding enables our model to accurately forecast extreme sea temperature anomalies of global key SST regions such as ENSO. Additionally, our model may have the ability to further predict ENSO-affected land regions across such as western North America and the Amazon. Our model achieves an 85\% accuracy hit rate in predicting ENSO events three months in advance. Our model has noticeably better skills than most ECMWF S2S models and the ensemble-averaged dynamics model, showing great potential.

In summary, we developed a novel deep learning model, CAS-Canglong, for predicting global sea surface temperature at each 0.25°×0.25° grid cell. Our validations and predictions demonstrate CAS-Canglong’s outstanding capability for predicting S2S SST changes. This model can be directly applied to other sub-seasonal Earth system weather prediction tasks, positioning it potentially to become potentially a generalizable standard deep learning model in geoscience.As SST serves as the foundation for S2S weather forecasts over land, this work lays the groundwork for land-based S2S weather predictions, such as precipitation, and has important implications for achieving skillful S2S forecasting in the future. 
\section{Method}
\subsection{Data}
We use the monthly aggregated ERA5 reanalysis data for atmospheric and oceanic variables for this study. The ERA5 dataset\textsuperscript{\cite{RN47}} represents the fifth-generation ECMWF atmospheric reanalysis of the global weather and climate, providing hourly reanalysis data spanning the past 60 years. As a cutting-edge generation of reanalysis data, ERA5 is recognized for its high quality by blending numerical models with reanalysis data using numerical assimilation methods. This approach has been used for numerous applications including global earth surface predictions\textsuperscript{\cite{RN48,RN49,RN50}}. 

We retained the highest resolution available from the ERA5 dataset, specifically 0.25°×0.25° at the Earth's surface, comprising a total of 1,440×721 grids for each variable at each time step. For predicting SST, we followed widely accepted practices with dynamical underpinning\textsuperscript{\cite{RN34,RN35.RN51}}, selecting single-level variables including sea level pressure, zonal and meridional wind speed and geopotential heights at 850-hPa and 500-hPa, surface latent heat flux, and net solar radiation as predictors. We adhered to a training-to-testing dataset division of 8:2. Our model was trained using monthly data from 1959 to 2009, while the testing phase utilized data from 2009 to 2022. 

Additionally, we also obtained SST data from another observations and modelled. In addition to the ERA5 SST data, the Hadley Centre Sea-Surface Temperature (HadSST) dataset is also used for model comparison as the observations. The HadSST dataset provides gridded monthly SST data from 1850 to the present. It fully incorporates \textit{in situ} measurements from ships and drifting buoys and is considered a high-quality, standard SST dataset\textsuperscript{\cite{RN52}}. Additionally, we utilized the European Centre for Medium-Range Weather Forecasts (ECMWF) S2S dataset and the North American Multi-Model Ensemble (NMME) S2S SST for comparison with CAS-Canglong. The NMME encompasses both real-time, initialized predictions and a comprehensive research database, archiving both retrospective and real-time forecasts\textsuperscript{\cite{RN53,RN54}}. For comparison, we employed NMME ensemble models that provide SST forecasts, including CanCM4i, Cansips, CanSIPS-IC3, CanSIPS-IC4, CanSIPSv2, CMC1-CanCM3, CMC2-CanCM4, COLA-RSMAS-CCSM3, COLA-RSMAS-CCSM4, COLA-RSMAS-CESM1, ECMWF, GEM-NEMO, GFDL-CM2p1-aer04, GFDL-CM2p5-FLOR-A06, GFDL-CM2p5-FLOR-B01, NASA-GEOSS2S, NASA-GMAO-062012, and NCAR-CESM1. The ECMWF S2S project and database include forecasts and reforecasts from both operational and research centers. We selected datasets with prediction lead times exceeding one month for comparison, including those from ECMWF, UKMO, HCMR, BoM, and CMA.

\subsection{CAS-Canglong model development}
Inspired by the successful applications of Transformer models\textsuperscript{\cite{RN30,RN42,RN55}} in spatiotemporal predictions and the unique challenges posed by SST-related ocean-atmosphere coupling and time-series modeling, we developed a computationally efficient Transformer-based model for SST predictions, named as ‘CAS-Canglong’. The acronym of CAS stands for the Chinese Academy of Sciences; Canglong is a figure from ancient Chinese mythology, believed to control rainfall.

This model is truly three-dimensional, designed to meet the complexities of sub-seasonal to seasonal scale forecasting. We incorporated time as the third dimension to consider predictors over multiple time steps. The model operates on an encoder-decoder framework utilizing the Transformer strategy. By employing the Swin-Transformer\textsuperscript{\cite{RN56}} architecture, it deeply extracts correlation information embedded in the predictors that is pertinent to land-atmosphere interactions. The data are processed through multi-head self-attention (MSA) and window multi-head self-attention modules (W-MSA), which upscale and restore the information to the output layer, enhancing prediction accuracy and detail\textsuperscript{\cite{RN57}}.

Specifically, this model is designed to handle normalized multivariable data in the format \textit{T\textsubscript{in }}× \textit{C }× \textit{N\textsubscript{lat }}× \textit{N\textsubscript{lon}}, where \textit{T\textsubscript{in}} represents the number of consecutive months used as input predictors (set to 16 months), \textit{C} is the number of features (set to 8 corresponding to the predictor variables), and \textit{N\textsubscript{lat} × N\textsubscript{lon}} are the latitude and longitude grid points (721 × 1440, respectively). Therefore, the total input dimension is 16 × 8 × 721 × 1440. The model outputs predictions for the subsequent months, with adjustments for various forecasting needs. Typically, predictions are made for the following three months with dimensions \textit{T\textsubscript{out }}× \textit{N\textsubscript{lat }}× \textit{N\textsubscript{lon }}= 3 × 721 ×1440, which is a standard setup for sub-seasonal scale predictions. The \textit{T\textsubscript{out}} parameter can be adjusted according to different prediction objectives, ranging from one to three, and up to a maximum of twelve months.

\begin{figure}[H] % picture
    \centering
    \includegraphics[width=1.0\columnwidth]{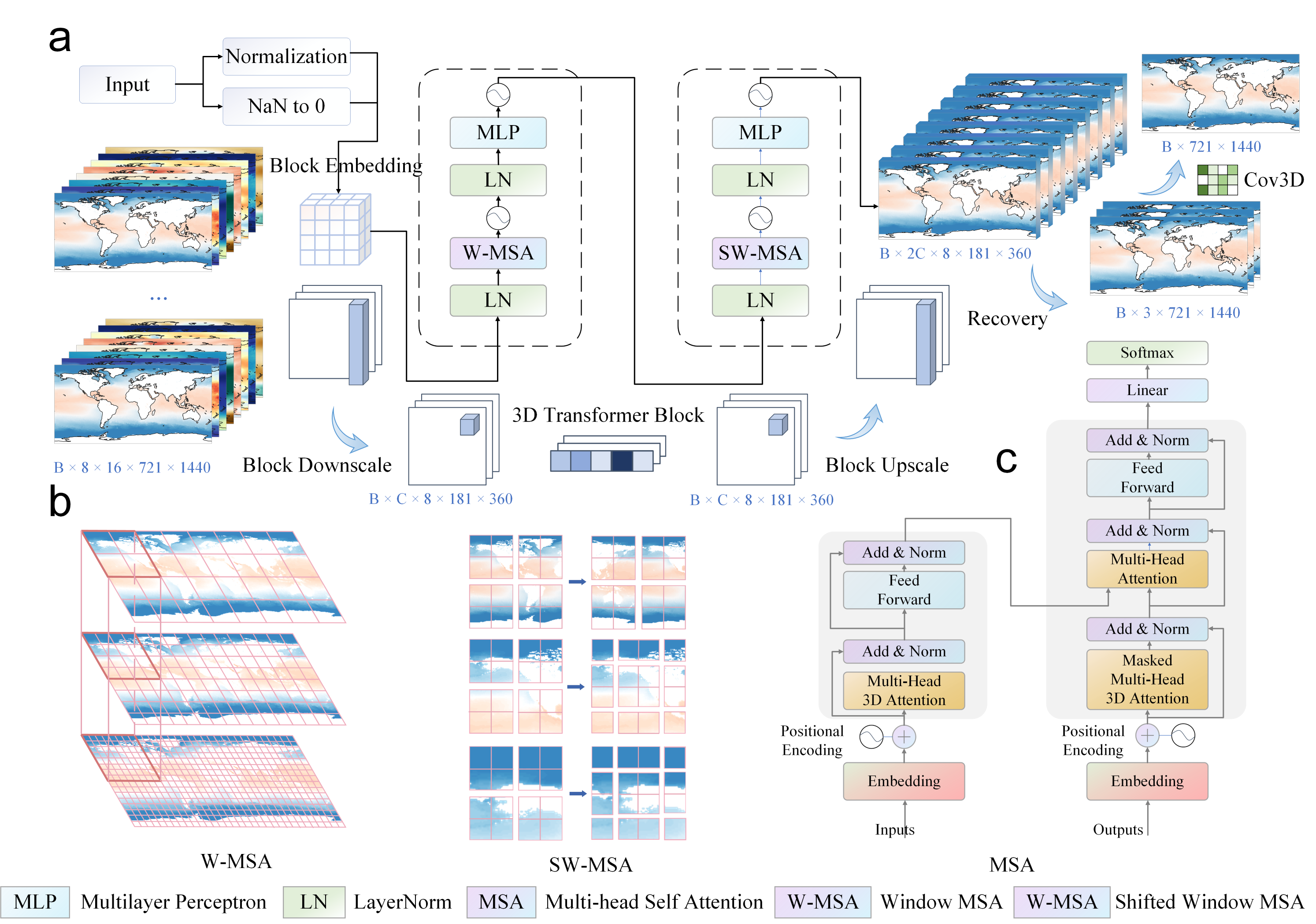}
    \caption{\textbf{Architecture of our deep learning model, CAS-Canglong. }(\textbf{a}) The architecture features a deep learning neural network which includes a data preprocessing and output recovery module. The core of the model is the Swin-Transformer, depicted in the center of the figure. The downscaled data enter the Swin-Transformer module, where features and relationships from the ocean-atmosphere system data are extracted through the MSA and W-MSA modules. (\textbf{b}) The schematic diagram illustrates the Swin-Transformer adapted for Earth surface tasks, with a progressively expanding receptive field to extract relationships within and between adjacent Earth pixels. (\textbf{c}) The Transformer module based on the encoder-decoder structure, highlighting the specific roles of MSA, followed by a linear output layer at the end of the decoder to finalize the prediction output.  
}
    \label{fig:fig5}
\end{figure}

As shown in Fig. \ref{fig:fig5}, the data undergo normalization and preprocessing before being passed through the W-MSA and SW-MSA processes, during which they are downscaled to extract features, inspired by the successful applications of Pangu-Weather and Fuxi-Weather models\textsuperscript{\cite{RN30,RN32}}. We employed a patch embedding layer with dimensions of 2 × 4 × 4, which is similar to that of the Vision Transformer. After this block embedding layer, the data are fed into the Swin Transformer, which utilizes an encoder-decoder architecture. To further tailor the model for Earth system applications, we implemented an Earth-specific positional bias strategy, which adapts techniques from computer vision tasks to Earth system imagery. This adaptation helps enhance the model's relevance and effectiveness in interpreting complex geographical and atmospheric data\textsuperscript{\cite{RN30}}. 

\subsection{Training approach}

The CAS-Canglong model processes batches of data (batch size B = 4), each containing a 16-month consecutive sequence and 8 features as multi-dimensional input. It predicts SSTs for several subsequent months as targeted outputs to train the model. Unlike other models that may use a rolling prediction strategy, this model focuses exclusively on high-precision SST simulations, intentionally disregarding other variables. Furthermore, it has the capability to dynamically adjust and output forecasts for up to 12 months in advance, accommodating a range of forecasting needs while maintaining a focus on accuracy. The period from January 1959 to August 2009 was used for training, from September 2009 to April 2016 for validation, and from May 2016 to December 2022 for testing. The ratio of the training, validation, and testing sets is approximately 8:1:1. 

The temporal period selection was carefully considered. Most of the validation and evaluation were conducted using dataset of validation and test period (September 2009 to December 2022). When comparing with other models, we selected the overlapping period for comparison. For instance, we selected the data starting from 2010 for NEMM comparison, and the data from 2015 for ECMWF S2S comparison.

To train the model efficiently and enhance prediction accuracy, we employed a combination of several loss functions. The primary metric utilized is the Mean Squared Error (MSE), which evaluates the model’s ability to accurately interpret SST data across a three-month observational period. This approach is critical in accelerating the model's convergence, as it facilitates a rapid and effective comprehension of the SST data, thereby enhancing the overall performance of the forecasting system: 
\begin{equation}
L_{fmse} = \frac{1}{T_{\text{out}}} \sum_{t=1}^{T_{\text{out}}} \left( X_{t,l} - \hat{X}_{t,l} \right)^2
\end{equation}

In this context,  $ X_{t,l} $ and  $\hat{X}_{t,l}$ represent the actual and predicted values of the SST for the month \textit{t + l}, respectively. The loss function \textit{L\textsubscript{fmse}} quantifies the model's capability to forecast up to \textit{T\textsubscript{out}} months lead time, and is a commonly utilized metric for temperature prediction. While MSE is effective in enhancing convergence by focusing on the model's ability to interpret SST data during the observation period, it may lead to a prioritization of short-term forecast accuracy over long-term precision. To address this potential imbalance, we further introduced a second loss function designed to distribute penalization more evenly across the entire forecast period. This strategic approach ensures a balanced focus on both immediate and extended forecast accuracy: 
\begin{equation}
L_{corr} = \frac{1}{T_{\text{out}}} \sum_{l=2}^{T_{\text{out}}} \text{argmax} \{ 0, 0.5 - \frac{\sum_{t=1}^{T} (x_{t,l} - \bar{X}_{l}) (\hat{x}_{t,l} - \bar{\hat{X}}_{l})}{\sqrt{\sum_{t=1}^{T} (x_{t,l} - \bar{X}_{l})^2} \sqrt{\sum_{t=1}^{T} (\hat{x}_{t,l} - \bar{\hat{X}}_{l})^2}} \}
\end{equation}
where  $\bar{X}_l$  and $\bar{\hat{X}}_l$  represent the mean observed and predicted values of sea surface temperature for the \textit{l}-th month following the observation period, respectively. The loss function \textit{L\textsubscript{corr}} specifically penalizes instances where the Pearson correlation coefficient between the actual and predicted SST values falls below 0.5 in long-term forecasts. This design underscores the critical importance of maintaining accuracy in long-term predictions, ensuring that the model not only predicts accurately in the short term but also retains reliability over extended periods. The combined loss function, defined as the sum of \textit{L\textsubscript{fmse}} and\textit{ L\textsubscript{corr}}, provides a comprehensive approach to model training, balancing the focus between immediate accuracy and long-term predictive reliability.

The Adam optimization algorithm is employed to optimize the model\textsuperscript{\cite{RN58}}, starting with an initial \textit{lr} = 0.0001 and using a cosine decay strategy for learning rate adjustment. This strategy gradually reduces the learning rate, aiding in fine-tuning the model's weights as training progresses. The model is initially trained on four A100 GPUs, each with 80GB of memory, which provides substantial computational power for handling the intensive calculations required for SST predictions. Additionally, the model is designed to support transfer learning, allowing for its adaptation to different SST regions. This flexibility enhances its utility across various geographic settings by fine-tuning the pre-trained model with data specific to each region. To further reduce prediction uncertainty, several identical models are trained using the same methodology but with different initialization parameters. This diversity in initial conditions enables ensemble predictions, where the outputs of multiple models are aggregated to produce a single forecast. This ensemble approach strongly enhances the reliability and accuracy of the forecasts by averaging out individual model biases and variances, providing a more robust prediction.

\subsection{SST regions and related climate indices}
Several observed SST and related indices across different regions of the world were used for model assessment: El Niño Southern Oscillation, Pacific Decadal Oscillation, Arctic Oscillation, Southern Oscillation and Indian Ocean Dipole. These regions are important for S2S forecasting. For instance, the SSTs in the South Atlantic, South Pacific, and Arctic play a crucial role in forecasting east Asian summer monsoon rainfall. These SSTs not only serve as key predictors but also enhance the timeliness of these forecasts, enabling predictions to be made up to three months in advance\textsuperscript{\cite{RN59}}.

The Niño 3.4 index is widely used to define the ENSO, a naturally occurring oceanic phenomenon involving atmospheric coupling in the tropical Pacific Ocean\textsuperscript{\cite{RN60,RN61}}. This index is based on the five-month rolling mean of the gridded SST in the Niño 3.4 region (170°W-120°W, 5°S-5°N). The two extreme phases of ENSO are determined by the SST anomaly persisting above or below 0.5°C for more than five months. El Niño is identified when the SST anomaly is above 0.5°C for more than five months, and La Niña is identified when the SST anomaly is below0.5°C for more than five months. 

The Pacific Decadal Oscillation (PDO) is defined as the leading pattern of the mean SST anomaly in the Pacific Ocean (110°E-100°W, 20°N-75°N). To calculate the leading pattern, the global mean SST anomaly for each month is first removed to minimize the influence of long-term trends in the data. A warm PDO phase corresponds to positive values, while a cold PDO phase corresponds to negative values\textsuperscript{\cite{RN62,RN63}}. 

The Artic Oscillation (AO) is a large-scale mode of climate variability. It is used to measure the 1000 mb height pressure poleward of 20°N. We assess the SST in this region\textsuperscript{\cite{RN64}}. 

The Southern Ocean (SO) is a dominant part of the global thermohaline circulation, playing a key role in climate variability and change\textsuperscript{\cite{RN65,RN66}}. We adopted the SO definition according to Luo et al\textsuperscript{\cite{RN67}}, which are divided into three regions: SAO (20°E-160°E, 40°S-60°S), SIO (160°E-110°W, 40°S-60°S), and SPO (20°E-160°E, 40°S-60°S). The Indo-Pacific Warm Pool (35°E-75°W, 15°S-15°N) is characterized by its SST consistently exceeding 28°C\textsuperscript{\cite{RN68}}. 

The Indian Ocean Dipole (IOD) is a climate phenomenon occurring in the Indian Ocean, defined by the SST difference between the eastern (90°E-110°E,10°S\~0°) and western regions (50°E-70°E,10°S-10°N) of the ocean\textsuperscript{\cite{RN69,RN70}}.
\subsection{Evaluation metrics}
We assessed two metrics, RMSE and correlation skill (R), defined as follows:
\begin{equation}
RMSE = \sqrt{\frac{\sum_{i=1}^{N_{\text{lon}}} \sum_{j=1}^{N_{\text{lat}}} W(i) \left( \hat{X}_{i,j,l} - X_{i,j,l} \right)^2}{N_{\text{lon}} \times N_{\text{lat}}}}
\end{equation}

\begin{equation}
R = \sqrt{\frac{\sum_{i=1}^{N_{\text{lon}}} \sum_{j=1}^{N_{\text{lat}}} W(i) \hat{X}_{i,j,l} X_{i,j,l}}{\sqrt{\sum_{i=1}^{N_{\text{lon}}} \sum_{j=1}^{N_{\text{lat}}} W(i) \hat{X}_{i,j,l}^2} \times \sqrt{\sum_{i=1}^{N_{\text{lon}}} \sum_{j=1}^{N_{\text{lat}}} W(i) X_{i,j,l}^2}}
}
\end{equation}
where $X_{i,j,l}$  and  $\hat{X}_{i,j,l}$ represent the ERA5 ground truth and predicted values of SST for the \textit{l} time month, respectively. \textit{W(i)} represents the area weighted formula of Earth, defined as follows:
\begin{equation}
W(i) = N_{\text{lat}} \times \frac{\cos \phi_i}{\sum_{i=1}^{N_{\text{lat}}} \cos \phi_i}
\end{equation}
where $\phi_i$ represents the latitude angle. The RMSE and R values were averaged over all times and weighted pixels. These two metrics are used to make comparison and evaluating prediction skills from different models in specific regions. For evaluating the CAS-Canglong model performance, we define the improvement as the RMSE reduction percentage from multi-model ensemble average to CAS-Canglong prediction. 

\section*{Acknowledgements}
\addcontentsline{toc}{section}{Acknowledgements}
This study was supported by the National Key R\&D Program of China (Grant No. 2022YFC3002804), the National Natural Science Foundation of China (Grant No. 42330506), and the Second Tibetan Plateau Scientific Expedition and Research Program (Grant nos. 2019QZKK0206 and 2019QZKK0208). X.Z. thanks the support from the Youth Innovation Promotion Association CAS (2022053) and the “Kezhen-Bingwei” Youth Talents Project from IGSNRR CAS (2021RC003). We extend our gratitude to the European Centre for Medium-Range Weather Forecasts (ECMWF) for providing the ERA5 dataset; the International Research Institute for Climate and Society (IRI) for the Niño 3.4 index ensemble prediction models; North American Multi-Model Ensemble (NMME) for the globally available S2S SST forecasts; Hadley Centre Sea Surface Temperature data set (HadSST) for the globally monthly SST for comparison. We appreciate the technical support from the National Supercomputer Center in Tianjin (NSCT). The work was also carried out at National Supercomputer Center in Tianjin. 

\section*{Data availability}
\addcontentsline{toc}{section}{Acknowledgements}
For training and testing, we downloaded a subset of the ERA5 dataset of 63 years from \href{https://cds.climate.copernicus.eu/}{https://cds.climate.copernicus.eu/}. Data used and produced in this research can be download from Figshare repository: doi.org/10.6084/m9.figshare.26779969. Supplementary figures could be downloaded from \href{https://github.com/GISWLH/CAS-Canglong}{https://github.com/GISWLH/CAS-Canglong}.

\section*{Code availability}
\addcontentsline{toc}{section}{Acknowledgements}
The code established based on Python PyTorch. We made use of the code base of Swin-transformer, available at \href{https://github.com/microsoft/Swin-Transformer}{https://github.com/microsoft/Swin-Transformer}. We released the trained models, inference code and the model details at a GitHub repository: \href{https://github.com/GISWLH/CAS-Canglong}{https://github.com/GISWLH/CAS-Canglong}.

\bibliographystyle{unsrt}  


\end{document}